\documentstyle[prl,aps,epsfig,multicol]{revtex}
\begin{document}
\tighten

\title{Fluctuation-Driven 1st-Order Isotropic-to-Tetrahedratic Phase 
Transition}
\author{Leo Radzihovsky}
\address{Department of Physics, University of Colorado, Boulder, CO 80309}
\author{T. C. Lubensky}
\address{Department of Physics, University of Pennsylvania, Philadelphia,
Pennsylvania 19174}

\date{\today}
\maketitle
\begin{abstract}

  Motivated in part by recent experiments on liquid crystals with
  bent-core molecules, which are observed to display a {\em
  spontaneous} chiral symmetry breaking, we introduce a field theory
  of a 3rd-rank tensor order parameter $T^{ijk}$ to describe the
  isotropic-to-tetrahedratic phase transition that we predict to take
  place in these materials. We study the critical properties of the
  corresponding phase transition and find that this transition,
  continuous at the mean-field level, is generically driven 1st-order
  by thermal fluctuations.

\end{abstract}
\pacs{PACS:}

\begin{multicols}{2}
  
Experimental studies of liquid crystals, systems that exhibit phases
that are intermediate (both in their properties and the symmetries
that they break) between ordinary liquids and solids continue to
present new surprises and theoretical challenges. The experimental
discovery in {\em achiral} bent-core molecules of
polar-ordered smectic layers by the Tokyo Tech group\cite{Watanabe}
and of spontaneous chiral symmetry breaking by the Boulder group\cite{Link} is
the latest important example of such a challenge. Bent-core
molecules are V-shaped with $C_{2v}$ symmetry\cite{Brand} characterized by a
non-polar direction ${\bf n}$ pointing from one endpoint of the V to
the other and an orthogonal polar direction ${\bf p}$ pointing to
the vertex of the V.  Seven distinct phases of bent-core molecules
tentatively labeled $B_1$ to $B_7$ have been
identified\cite{Bphases}, though not all have been fully
characterized.  Two phases, the $B_2$ and $B_7$ phases, are smectic
phases consisting of stacks of fluid layers with some internal tilt
order.  The Boulder group\cite{Link} has shown that the $B_2$ phase
is an anti-ferrolectric smectic-$C$ phase in which the configuration
of molecules in each layer is chiral with chirality alternating in
sign from one layer to the next. A material of {\em achiral} nematogens 
having a ground state which is ferroelectric and {\em homogenously} 
chiral has also recently been discovered.\cite{Walba}  
Within smectic phases, such {\em spontaneous} breaking of chiral symmetry is
possible because layers provide an anisotopic environment, 
defined by the layer normal, relative to which molecules can tilt and
rotate.  

Though most bent-core molecules do not exhibit nematic phases, some
do\cite{bananaNem}, and it is of some interest to develop the minimal
theory capable of exhibiting isotropic, nematic, and the various
smectic phases of bent-core molecules.  Homogeneous (i.e.,
translationally-invariant) phases composed of achiral molecules and
characterized only by {\em two} orthogonal axes cannot be chiral.
Clearly, then, a theory, whose orientational order is based only on
the Maier-Saupe symmetric-traceless 2nd-rank tensor order parameter
$Q_{ij}$ cannot describe the exotic orientational order of the $B_2$
phase.  In many liquid crystals, electric dipoles play an
insignificant role in determining the development of order, and it is
appropriate to consider models in which molecular properties are
described entirely by their mass-moment tensors. Since the first
mass-moment tensor relative to the center of mass is zero, these
molecules do not have mass-moment {\em vector} order parameter.  They
do, however, have a {\em third}-rank mass-moment tensor that does
contain information about the polar shape of a bent-core molecule.
The simplest model capable of describing both the nematic and the
orientational order of the aforementioned smectic phases of bent-core
molecules is one with both 2nd-rank and 3rd-rank order parameters.
Arbitrary symmetric third rank tensors can be decomposed into a
symmetric-traceless part $T^{ijk}$, which transforms under the $l=3$
representation of the group $SO(3)$, and a part that transforms like a
vector $p^i$ ($l=1$) under $SO(3)$.

A model with $Q^{ij}$, $T^{ijk}$ and $p^i$ order parameters produces
an incredibly rich phase diagram at the mean-field level.\cite{LR} It
predicts, in particular, translationally invariant nematic phases with
$D_{3h}$, $D_{2d}$ and even chiral $D_2$ symmetry in addition to the
usual uniaxial and biaxial nematic phases with respective $D_{\infty
  h}$ and $D_{2h}$ symmetry.  The latter phases are characterized by
$p^i=0$, $T^{ijk}=0$, and $Q^{ij}\neq 0$, whereas the former, more
unfamiliar phases have both $T^{ijk}$ and $Q^{ij}$ nonzero, but
$p^i=0$. There is also a novel tetrahedratic (T) phase with $p^i=0$,
$Q^{ij}=0$, and only $T^{ijk}$ nonzero. The {\em direct} transition
from the isotropic phase to the tetrahedratic phase\cite{Fel} can be
described by a simpler model depending only on the 3rd-rank tensor
order parameter $T^{ijk}$.  Since there are no odd order invariants of
$T^{ijk}$, this transition can be continuous in mean-field
theory.\cite{LR} In this note, we study fluctuation corrections to the
mean-field isotropic-to-tetrahedratic (IT) transition within an
$\epsilon$-expansion about the upper critical dimension $d_c =4$. Our
principal result is that the presence of two distinct fourth order
invariants (see Eq.\ (\ref{Ht}) below) leads to a runaway within the
$\epsilon$-expansion whose effect in analogy with charged
superconductors\cite{HLM} and a ferromagnet in a cubic
field\cite{Rudnick} is to convert this mean-field 2nd-order transition
to a 1st-order one.

Our starting point is the Landau free-energy density functional ${\cal
  H}[T_{ijk}]$, controlling the {\em direct} transition into the
tetrahedratic phase:
\begin{eqnarray}
{\cal H}&=&{1\over2}(\partial_l T^{ijk}\partial_l T^{ijk})+
{1\over2} r T^{ijk} T^{ijk}\nonumber\\
&+& u(T^{ijk} T^{ijk})^2 + v T^{i_1 i_2 i_3} T^{i_1 i_4 i_5}
T^{i_2 i_4 i_6} T^{i_3 i_5 i_6}\;.
\label{Ht}
\end{eqnarray}
In the above expressions we have used the Einstein summation
convention, and have left out the dipolar-like gradient terms
$\partial_{i_1} T_{i_1 j k}\partial_{i_2}T_{i_2 j k}$ that couple
internal indices of $T_{ijk}$ to that of spatial coordinate $\bf x$.
This latter simplification, quite commonly employed in the analysis of
the isotropic-nematic (IN) transition, might in principle modify the
nature of the IT transitions at the longest scales, as it does for
example in ferromagnets.\cite{dipolarFM} We leave the analysis of
these ``dipolar'' effects to a future investigation. We have also
omitted couplings of $T^{ijk}$ to other order parameters.  Since here
we are interested in the {\em direct} transition to the T phase, all
other order parameters (e.g., the nematic $Q_{ij}$) are generically
noncritical and can, therefore, be safely integrated out, only
finitely renormalizing the Landau parameters in ${\cal H}$.  As usual
the quadratic parameter $r\sim T - T_c$ vanishes at the mean-field
transition temperatures $T_c$, which is determined predominantly by
the interaction potential between molecules.  In writing down quartic
nonlinearities in ${\cal H}$ we have used a non-obvious relation,
\begin{eqnarray}
{1\over2}(T^{ijk} T^{ijk})^2&=&T^{i j_1 k_1}T^{l j_1 k_1} T^{i j_2
k_2} T^{l j_2 k_2}\nonumber\\
&+&T^{i_1 i_2 i_3} T^{i_1 i_4 i_5} T^{i_2 i_4 i_6}T^{i_3 i_5 i_6}\;,
\label{quartics}
\end{eqnarray}
valid when the dimension $d_T$ of the space in which $T^{ijk}$ rotates
is three, to reduce the number of independent quartic couplings in
${\cal H}[T]$ from three to two. The coupling $v$ was incorrectly
omitted in the analysis of Fel.\cite{Fel} If $v=0$, the model
described by Eq.\ (\ref{Ht}) has $O(7)$ symmetry. Thus, it is
analogous to an $O(N)$ Heisenberg ferromagnet with an $N$-dimensional
vector order parameter $T_\alpha$ in a cubic crystal field, giving rise
to a $v\sum_\alpha T_\alpha^4$ quartic potential in addition to the
$O(N)$ invariant potential $u(\sum_\alpha T_\alpha^2)^2$.

As discussed extensively in Ref.\onlinecite{LR}, the IT transition is
generically continuous at the mean-field level since rotational
invariance forbids the appearance of a cubic invariant in $T^{ijk}$.
Its nature, however, depends on the sign of $v$. For $v>0$, the stable
low-temperature phase has $D_{3h}$ symmetry, whereas for $v<0$, it has
tetrahedral ($T_d$) symmetry. Similar behavior is seen in $O(N)$
ferromagnets in a cubic crystal field.  When the cubic coupling $v$ is
nonzero, the low-temperature phase has $T_\alpha\sim (1,0,...,0)$ for
$v<0$ and $T_\alpha\sim (1,1,...,1)/\sqrt{N}$ for $v>0$.  Our main goal
then is to go beyond this mean-field theory\cite{LR} and study effects
of thermal fluctuations on this heretofore unexplored IT transition.

Before analyzing fluctuations in this model, it is useful to introduce
a tensor representation for the 7 independent components of $T^{ijk}$.
Any third-rank symmetric-traceless field $T^{ijk} ( {\bf x})$ can be
expressed as $T^{ijk} ( {\bf x}) = \sum_{\alpha=1}^7 T_\alpha( {\bf x}
) I_\alpha^{ijk}$, where
\begin{mathletters}
\begin{eqnarray}
I_1^{ijk}& = &\sqrt{5/2}[n^i n^j n^k - {1 \over 5} (\delta^{ij} n^k +
\delta^{jk} n^i +\delta^{ki} n^j)]\\
I_2^{ijk} &=& {1 \over 2} (m^i m^j m^k - m^i l^j l^k - m^j l^k l^i -
m^k l^i l^j ) \\
I_3^{ijk} & = & {1 \over 2} (l^i l^j l^k - i^i m^j m^k - l^j m^k m^i
- l^k m^i m^j ) \\
I_4^{ijk}& = & \sqrt{5/2}[m^i n^jn^k + m^j n^k n^i + m^k n^i
n^j\nonumber\\
&-&{1\over 5}(m^i\delta^{jk}+m^j \delta^{ik} + m^k \delta^{ij})]\\
I_5^{ijk} & = & \sqrt{5/2}[l^i n^j n^k + l^j n^k n^i + l^k n^i
n^j\nonumber\\
&-&{1
\over 5}(l^i\delta^{jk}+l^j \delta^{ik} + l^k \delta^{ij})]\\
I_6^{ijk} & = & {1\over \sqrt{6}}[n^i (m^j m^k - l^j l^k) + n^j (m^im^k -
l^i l^k)\nonumber\\
&+& n^k (m^i m^j - l^i l^j)]\\
I_7^{ijk} & = & {1\over\sqrt{6}}(n^i m^j l^k + n^i l^j m^k + m^i l^j n^k +
m^i n^j l^k \nonumber\\
&+& l^i n^j m^k + l^i m^j n^k)
\end{eqnarray}
\label{Is}
\end{mathletters}
are an orthonormal set of basis tensors with $n^i$, $m^i$, and $l^i$
forming a space-fixed right-handed orthonormal basis.  The $O(7)$
invariance of the Eq.\ (\ref{Ht}) when $v=0$ follows easily from the
relation $T^{ijk} T^{ijk} = \sum_{\alpha=1}^7 T_\alpha^2$.

The effective Hamiltonian $H=\int d^d x{\cal H}$ in $d$ spatial
dimensions can be conveniently rewritten in terms of Fourier
transformed $T$-fields
\begin{eqnarray}
H&=&{1\over2}\int_{\bf k}(k^2+r)|T_{ijk}({\bf k})|^2\nonumber\\
&+&\int_{\bf x}V_{i_1 j_1 k_1,i_2 j_2 k_2}^{i_3 j_3 k_3,i_4 j_4 k_4}
T_{i_1 j_1 k_1} T_{i_2 j_2 k_2} T_{i_3 j_3 k_3} T_{i_4 j_4 k_4}\;,
\label{H}
\end{eqnarray}
where $V_{i_1 j_1 k_1,i_2 j_2 k_2}^{i_3 j_3 k_3,i_4 j_4 k_4}\equiv
V_{{\bf i}_1,{\bf i}_2}^{{\bf i}_3,{\bf i}_4}$ is a fully symmetrized
quartic potential (consisting of products of 3 Kronecker
$\delta$-functions), containing the $u$ and $v$ vertex couplings and
is too complicated to be reproduced here. There are at least two ways
to generalize our model, originally defined in $3$ dimensions to
arbitrary spatial dimension $d$.  We can either fix the dimension
$d_T$ of the space in which $T^{ijk}$ ``lives'' to be three and let
the dimension $d$ of space be arbitrary, or we could lock $d_T$ and
$d$ together.  Here, we adopt the former option, keeping $d_T = 3$.
If we had included dipolar-like gradient coupling terms in Eq.\ 
(\ref{Ht}), we would have had no choice but to use the analytical
continuation in which $d_T = d$. The advantage of symmetrized
representation in Eq.\ (\ref{H}) is that averages of any pair of
$T^{ijk}$ fields appearing in the quartic operator are equivalent, and
the field theory part of the calculation is effectively mapped onto a
standard analysis of a scalar $\phi^4$ field theory.

It is not difficult to show that a corresponding tensor part of the
$T-T$ propagator $G_{ijk}^{lmn}$ defined by
\begin{equation}
\langle T_{i j k}({\bf k}) T_{l m n}({\bf k}')\rangle
=(2\pi)^d \delta^{(d)}({\bf k}+{\bf k}'){G_{i j k}^{l m n}\over
k^2+r}
\label{TT}
\end{equation}
is given by
\begin{eqnarray}
6 G_{i j k}^{l m n}&=&\big[\delta_{il}\delta_{jm}\delta_{kn}+5 \mbox{
permutations of}\;\; (i j k)\big]\nonumber\\
&-&{2\over5}\big[\delta_{ij}(\delta_{lm}\delta_{kn}+
\delta_{lk}\delta_{mn}+\delta_{ln}\delta_{km})\nonumber\\
&+&(i\leftrightarrow k)+(i\leftrightarrow j)\big]
\label{G}
\end{eqnarray}
The tensor $G_{i j k}^{l m n}$ is uniquely constrained by the
rotational invariance to be a sum of products of Kronecker
$\delta$-functions, with two nontrivial coefficients uniquely
determined by the various traces of $G_{i j k}^{l m n}$. Another more
systematic way of obtaining $G_{i j k}^{l m n}$ is to use the $O(7)$
representation of $T_{i j k}({\bf k})$, Eq.\ (\ref{Is}) inside the
left hand side of Eq.\ (\ref{TT}) to produce
\begin{equation}
\hspace{-0.5cm}\langle T_{i j k}({\bf k}) T_{l m n}({\bf
k}')\rangle =\sum_{\alpha,\alpha'=1}^7 I_\alpha^{ijk}
I_{\alpha'}^{lmn}\langle T_\alpha({\bf k}) T_{\alpha'}({\bf  k}')\rangle\;. 
\label{TTO7}
\end{equation}
Using the $O(7)$ representation of $H$ along with equipartition,
we obtain
%
%
$\langle T_\alpha({\bf k}) T_{\alpha'}({\bf k}')\rangle =(2\pi)^d
\delta^{(d)}({\bf k}+{\bf k}'){\delta_{\alpha\alpha'}/(k^2+r)}$, which when
used inside Eq.\ (\ref{TTO7}) gives
%
%
$G_{i j k}^{l m n}=\sum_{\alpha=1}^7 I_\alpha^{ijk} I_{\alpha}^{lmn}$
that can be shown to be equivalent to our expression for $G_{i j k}^{l
  m n}$ in Eq.\ (\ref{G}).

To assess the role of thermal fluctuations, we need to compute the
total free energy by integrating over the field $T^{ijk}({\bf
x})$ rather than by simply minimizing the effective Hamiltonian
above, as we did in Ref.\cite{LR}. As usual, because of the
nonlinear quartic interaction this cannot be done exactly, and
near a critical point ($r\approx0$), a perturbative calculation
in $V_{{\bf i}_1,{\bf i}_2}^{{\bf i}_3,{\bf i}_4}$ diverges (for
$d<4$), thereby indicating that the nature of the IT transition
is qualitatively modified by thermal fluctuations.

To make sense of these divergences we employ the standard perturbative
momentum-shell RG calculation\cite{Wilson}, in which we integrate out
an infinitesimal fraction (a momentum shell) of degrees of freedom at
a time near the shortest length scales, at the lattice cutoff
$\Lambda^{-1}$. More concretely, we separate the $T_{ijk}$ order
parameter field into high and low-wavevector components $T_{ijk}({\bf
  x})=T_{ijk}^<({\bf x})+T_{ijk}^>({\bf x})$, with $T_{ijk}^>$ having
support in the wavevector range $\Lambda e^{-\ell}<k<\Lambda$ and
$\Lambda$ is an ultraviolet (uv) cutoff of order of the inverse of
molecular size. We integrate out the high-wavevector part
$T_{ijk}^>({\bf x})$, perturbatively in $V_{{\bf i}_1,{\bf i}_2}^{{\bf
    i}_3,{\bf i}_4}$, and rescale the lengths and long-wavelength part
of the fields according to ${\bf x}=e^{\ell}{\bf x'}$, $T_{ijk}^<({\bf
  x})=e^{\ell(2-d)/2}T_{ijk}({\bf x'})$, so as to restore the uv
cutoff back to $\Lambda$ and to (1-loop order) keep the coefficient of
the $k^2 |T_{ijk}(k)|^2$ in $H$ fixed at $1$.  Under this
transformation the resulting effective free energy functional $H$ can
be restored into its original form Eq.\ (\ref{H}) with effective
$\ell$-dependent couplings.
\columnwidth3.4in
\begin{figure}[bth]
\centering
\setlength{\unitlength}{1mm}
\begin{picture}(40,40)(0,0)
\put(-45,-85){\begin{picture}(0,0)(0,0)
\includegraphics{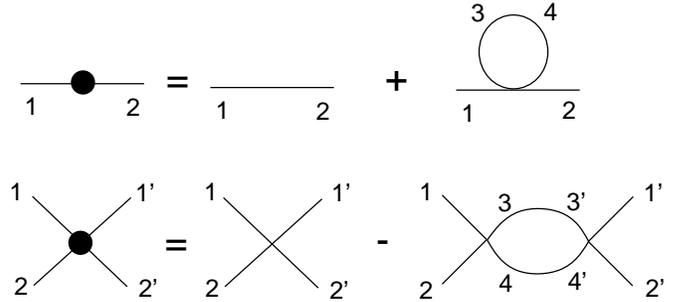}
\end{picture}}
\end{picture}
\caption{Graphical representation of the leading order correction
  to the reduced temperature $r$ and the quartic coupling $V_{{\bf
      i}_1,{\bf i}_2}^{{\bf i}_3,{\bf i}_4}$.}
\label{delta_rV}
\end{figure}

Standard perturbative 1-loop RG analysis, represented graphically in
Fig.\ref{delta_rV} leads to the following flow equations
\begin{eqnarray}
\hspace{-0.5cm}{d\over d\ell}G^{-1}_{{\bf i}_1,{\bf i}_2}&=&
2G^{-1}_{{\bf i}_1,{\bf i}_2}
+{12\Lambda^{d-2}C_d\over(1+r/\Lambda^2)}\;
V_{{\bf i}_1,{\bf i}_2}^{{\bf i}_3,{\bf i}_4}\;
G_{{\bf i}_3,{\bf i}_4}\;,\label{dV}
\\
{d\over d\ell}V_{{\bf i}_1,{\bf i}_2}^{{\bf i}_1',{\bf i}_2'}&=&
\epsilon\; V_{{\bf i}_1,{\bf i}_2}^{{\bf i}_1',{\bf i}_2'}
\nonumber\\
& & -{36\Lambda^{d-4}C_d\over(1+r/\Lambda^2)^2}\; V_{{\bf
i}_1,{\bf i}_2}^{{\bf i}_3,{\bf i}_4}\; V_{{\bf i}_1',{\bf
i}_2'}^{{\bf i}_3',{\bf i}_4'}\; G_{{\bf i}_3,{\bf i}_3'}G_{{\bf
i}_4,{\bf i}_4'}\;,\label{dV2}
\end{eqnarray}
where $\epsilon = 4 -d$.  These equations describe the evolution of
the reduced temperature $r(\ell)$ (determined by the $q=0$ part of the
2-point inverse propagator $G^{-1}_{{\bf i}_1,{\bf i}_2}(q=0,r)$) and
the quartic vertex $V_{{\bf i}_1,{\bf i}_2}^{{\bf i}_3,{\bf
    i}_4}(\ell)$ after a fraction $e^{-\ell}$ of high-wavevector modes
has been integrated out.  In above
$C_d=2\pi^{d/2}/\left((2\pi)^d\Gamma(d/2)\right)$ is a surface area of
a $d$-dimensional sphere divided by $(2\pi)^d$, an obvious extension
of Einstein convention was employed, and $\epsilon$ was taken to be
small so as to allow higher-order corrections in $V_{{\bf i}_1,{\bf
    i}_2}^{{\bf i}_3,{\bf i}_4}$ to be neglected. We note in passing
that for the fully symmetrized $V_{{\bf i}_1,{\bf i}_2}^{{\bf
    i}_3,{\bf i}_4}$ vertex, the factors $12$ and $36$ of the
respective second term in Eqs.\ (\ref{dV}) and (\ref{dV2}) are
identical to the corresponding coefficients in the equations for the
reduced temperature and quartic coupling constant in the scalar
$\phi^4$ (Ising) model.\cite{Wilson}

After technically tedious but conceptually straightforward tensor
contraction in Eq.\ (\ref{dV2}), we find that the RG flow
equations at criticality ($r=0$) for the two dimensionless
couplings of our theory $\tilde{u}\equiv u\Lambda^{d-4}C_d$ and
$\tilde{v}\equiv v\Lambda^{4-d}C_d$, with $u$ and $v$ defined in
Eq.\ (\ref{Ht}), are
\begin{mathletters}
\begin{eqnarray}
{d\tilde{u}(\ell)\over d\ell}&=&\epsilon \tilde{u} - 60
\tilde{u}^2 - {28\over5} \tilde{u} \tilde{v} -
{17\over25} \tilde{v}^2\;,\label{ut}\\
{d\tilde{v}(\ell)\over d\ell}&=&\tilde{v}(\epsilon - {116\over25}
\tilde{v} - 48 \tilde{u})\;.\label{vt}
\end{eqnarray}
\label{uv}
\end{mathletters}
The flow off the critical surface, controlled by $\tilde{r}(\ell)\equiv
r(\ell)/\Lambda^2$ is determined by the equation
\begin{equation}
{d\tilde{r}(\ell)\over d\ell}=2\tilde{r}
+{1\over1+ \tilde{r}}(36\tilde{u} +{76\over5}\tilde{v})\;.\label{rt}
\end{equation}
As a simple check on our results, we first note that the flow for
$\tilde{v}$, Eq.\ (\ref{vt}) has an obvious fixed point solution
of $\tilde{v}=0$. This property of the flow for $\tilde{v}$ is a
rigorous consequence of an exact $O(7)$ invariance of our theory
for $\tilde{v}=0$, as can be easily seen from Eq.\ (\ref{Ht}).

Standard fixed point analysis of the above equations leads to two
physical critical points given by: (i) Gaussian with
$\tilde{u}=0,\tilde{v}=0$, and (ii) O(7) with
$\tilde{u}={\epsilon/60},\tilde{v}=0$. As a further check, we also
note that for $v=0$, above flow equations, Eqs.\ (\ref{rt}) and
(\ref{ut}) reduce exactly to the flow equations for the $O(N)$
Heisenberg model with $N=7$ (the model that our Hamiltonian $H$
explicitly reduces to for $v=0$), with corresponding universal
coefficients $36$ and $60$, in above equations, identical to the
well-known $4(N+2)$ and $4(N+8)$ coefficient for the $O(N)$ Heisenberg
model with $N=7$.

As expected and discussed above, for $d<4$ the Gaussian critical point
is unstable. For the special $v=0$ $O(7)$ invariant subspace, this
instability terminates at the well-known $N=7$ Heisenberg model
critical point characterized by (1-loop) exponents:
$\nu=1/(2-3\epsilon/5)$, $\gamma=2\nu$ and $\eta=0$.
\columnwidth3.4in
\begin{figure}[bth]
\centering
\setlength{\unitlength}{1mm}
\begin{picture}(70,70)(0,0)
\put(-30,-40){\begin{picture}(30,30)(0,0)
\includegraphics{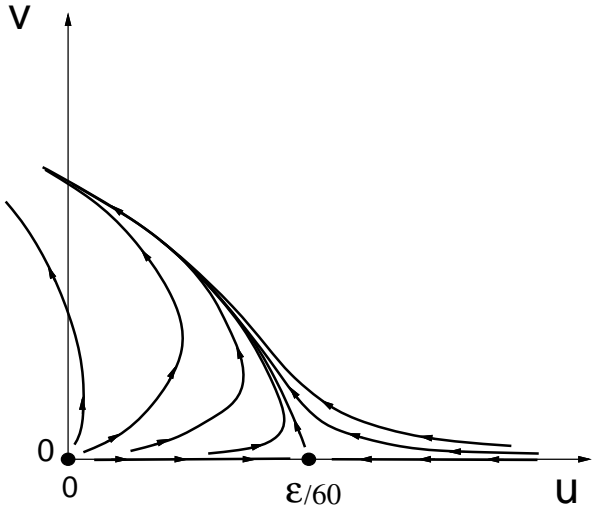}
\end{picture}}
\end{picture}
\caption{RG flow characterizing the IT phase transition.
  The Gaussian and $O(7)$ ($u=\epsilon/60,v=0$) critical points are
  indicated by black circles. In a generic model, with $v\neq0$, the
  flow runs away into region of $u<0$, suggesting a fluctuation-driven
  1st-order IT transition.}
\label{flow}
\end{figure}
This $O(7)$ critical point is also unstable with respect to
turning on $v$. Standard linear analysis of Eqs.\ (\ref{uv})
around the $O(7)$ ($v=0$) critical point leads to two
eigenvectors and eigenvalues that characterize the instability to
turning on the $O(7)$-symmetry breaking perturbation $v$
\begin{mathletters}
\begin{eqnarray}
(\delta u,\;\;\delta v)_u
&=&\tilde{u}(1\;,\;\;0)\;,\;\;\;\lambda_{\tilde{u}}=-\epsilon\;,
\label{delta_u}\\
(\delta u,\;\;\delta v)_v
&=&\tilde{v}(-{7\over90}\;,\;\;1)\;,\;\;\;\lambda_{\tilde{v}}
={\epsilon\over5}\;.\label{delta_v}
\end{eqnarray}
\end{mathletters}
As is graphically displayed in Fig.\ref{flow} for the critical surface
$r_{c}(\ell)$, we find that in the full model the effective couplings
$u(\ell)$ and $v(\ell)$ run off to infinity.

The absence of a stable fixed point and the runaway towards
negative values of $u$ is analogous to the situation encountered
in model superconductors\cite{HLM} with $(N/2)$-component $U(1)$
symmetry in which there is runaway from an $O(N)$-symmetric
Heisenberg fixed point towards negative $u$ when charge is turned
on. In the latter system, the runaway can be identified with a
fluctuation induced first-order phase transition\cite{HLM,CLN}.
It is likely that the runaway in the current problem is also a
signal of a fluctuation-induced first-order IT transition.  In the
case of the superconductor, an effective free energy indicating a
first-order transition could be obtained by integrating out the
vector potential.  In the model considered here, there is no
obvious analog of the vector potential.  As a result, the
detailed analysis of the first-order IT transition is likely to be
more similar to that of the fluctuation induced first-order
transition, which exists for appropriate values of the
potentials, in $O(N)$ magnets in a cubic field\cite{Brezin},
analyzed for $N=2$ by Rudnick\cite{Rudnick}.

Leo Radzihovsky acknowledges support by the NSF through the CAREER
grant DMR96-25111, the MRSEC grant DMR98-09555, and by the A. P.
Sloan and David and Lucile Packard Foundations.  Tom Lubensky was
supported by the NSF through grant DMR97-30405.

\end{multicols}

\begin{references}
\vspace{-1cm}
\bibitem{Watanabe} T. Niori, {\it et al.},
J. Mater. Chem. {\bf 6}, 1231 (1996).
\bibitem{Link} D. R. Link, {\it et al.}, Science {\bf 278}, 1924 (1997).
\bibitem{Brand} H.R. Brand, {\it et al.},
Europhys. J. B {\bf 6}, 347 (1998).
\bibitem{Bphases} G. Pelzl, S. Diele, W. Weissflog, Adv. Mater. {\bf 11},
707 (1999).
\bibitem{Walba} D. Walba, {\it et al.}, Science {\bf 288}, 2181 (2000). 
\bibitem{bananaNem} J. Matraszek, {\it et al.}, Liq. Crystals {\bf 27} 429
(2000); K. Kishikawa, {\it et. al.} Chem. Matt. {\bf 11}, 867
(1999).
\bibitem{LR} T. C. Lubensky and L. Radzihovsky, unpublished.
\bibitem{Fel} see e.g., L. G. Fel, Phys. Rev. E {\bf 52} 702 (1995).
\bibitem{HLM} B.I. Halperin, T.C. Lubensky, and S. K. Ma,
Phys. Rev. Lett. {\bf 32}, 292 (1974)
\bibitem{Rudnick} J. Rudnick, Phys. Rev. B, {\bf 18}, 1406 (1978).

\bibitem{dipolarFM} A. Aharony, M. E. Fisher, Phys. Rev. B {\bf 8},
  3323 (1973).
\bibitem{Wilson}  K. G. Wilson and J. Kogut, Phys. Rep. C, {\bf 12}, 77
(1977).
\bibitem{CLN} J.-H. Chen, T.C. Lubensky, and D.R. Nelson,
Phys, Rev. B {\bf 17}, 4274 (1978).
\bibitem{Brezin} E. Brezin, LeGuillou and J. Zinn-Justin,
Phys. Rev. B, {\bf 10}, 892 (1974).

\end{references}
\end{document}